*Article*

# Major Contribution of Halogenated Greenhouse Gases to Global Surface Temperature Change

**Qing-Bin Lu**

Department of Physics and Astronomy and Departments of Biology and Chemistry, University of Waterloo, 200 University Avenue West, Waterloo N2L 3G1, ON, Canada; qblu@uwaterloo.ca

**Abstract:** This paper aims to better understand why there was a global warming pause in 2000–2015 and why the global mean surface temperature (GMST) has risen again in recent years. We present and statistically analyze substantial time-series observed datasets of global lower-stratospheric temperature (GLST), troposphere–stratosphere temperature climatology, global land surface air temperature, GMST, sea ice extent (SIE) and snow cover extent (SCE), combined with modeled calculations of GLSTs and GMSTs. The observed and analyzed results show that GLST/SCE has stabilized since the mid-1990s with no significant change over the past two and a half decades. Upper-stratospheric warming at high latitudes has been observed and GMST or global land surface air temperature has reached a plateau since the mid-2000s with the removal of natural effects. In marked contrast, continued drastic warmings at the coasts of polar regions (particularly Russia and Alaska) are observed and well explained by the sea-ice-loss warming amplification mechanism. The calculated GMSTs by the parameter-free quantum-physics warming model of halogenated greenhouse gases (GHGs) show excellent agreement with the observed GMSTs after the natural El Niño southern oscillation and volcanic effects are removed. These results have provided strong evidence for the dominant warming mechanism of anthropogenic halogenated GHGs. The results also call for closer scrutiny of the assumptions made in current climate models.

**Citation:** Lu, Q.-B. Major Contribution of Halogenated Greenhouse Gases to Global Surface Temperature Change. *Atmosphere* **2022**, *13*, x. https://doi.org/10.3390/xxxxx

Academic Editor(s): Katerina Pantavou and Kostas Philippopoulos

Received: 3 August 2022
Accepted: 29 August 2022
Published: date

**Publisher's Note:** MDPI stays neutral with regard to jurisdictional claims in published maps and institutional affiliations.

## 1. Introduction

It is generally agreed that the measured global mean surface temperature (GMST) had a small rise of 0.2–0.3 K between 1850–1900 and 1950–1970, which was mainly due to natural forces [1-5], and a drastic rise of approximately 0.6 K between 1950–1970 and around 2000 [5-8]. A vast amount of literature has documented a global warming pause for the period between 2000 and around 2015 [4-6, 9-20], whereas the GMST appears to have risen again in recent years [8]. El Niño southern oscillation (ENSO) is one of the largest sources of year-to-year variability, but it is not likely to be the main cause of the changes in GMST since 1950–1970 [21,22]. This paper is devoted to understanding global climate change since the 1970s when the emission of anthropogenic halogenated gases (mainly chlorofluorocarbons—CFCs) into the atmosphere became significant.

In 1975, Ramanathan [23] made the first theoretical work showing that CFCs are much more potent greenhouse gases (GHGs) than $CO_2$ on a molecule-by-molecule basis. Subsequently, researchers [24-27] have shown substantial contributions of halogenated ozone-depleting substances (ODSs) to global warming or Arctic warming under climate models, assuming $CO_2$ as the major culprit of warming, while a CFC-warming model of GMST has been proposed in a series of previous publications by the present author [4,5,13,14], showing the dominant warming effect of halogenated GHGs since the 1970s.

The global cooling in lower-stratospheric temperature from the late 1970s to the mid-1990s is well known to have been caused by ODSs, as ozone is the main solar radiation absorber therein. It has been widely documented that ODSs have played the dominant role in controlling the lower-stratospheric temperature since 1979 [7, 28, 29]. These studies found that for the period 1979–2014, the response of ozone and lower-stratospheric temperature is detectable in the observations to ODSs but not well-mixed (non-halogenated) GHGs and that the cooling trends in global and tropical lower-stratospheric temperatures have ceased since around 1997, consistent with the trends of stratospheric ODSs, which peaked around 1997. It is worth noting that increased levels of unsaturated GHGs should lead to a significant direct radiative cooling in the upper stratosphere, particularly at altitudes above 25 km, accompanying a warming on the ground, but this cooling is negligibly small in the lower stratosphere [30, 31]. It has also been noted that global and tropical lower-stratospheric temperatures have exhibited nonmonotonic and nonlinear trends and such complexities have offered challenges to understanding the underlying causes [29].

In recent papers [32, 33], the present author reported the fingerprints of the cosmic-ray-driven electron reaction (CRE) mechanism in the Antarctic ozone hole [34-36] and the discovery of a large, deep and all-season tropical ozone hole, respectively. It has been robustly demonstrated that both ozone and lower-stratospheric temperature have been dominantly governed by not only ODSs but cosmic-ray (CR) intensity [4, 5, 13, 32, 33]. It was shown that after removal of the CR effect, the Antarctic ozone hole has shown a pronounced recovery since the mid-1990s [4, 5, 32], closely following the change trend of anthropogenic ODSs due to the Montreal Protocol [7]. A similar conclusion was also reached by others [37, 38]. It is reasonable to expect a similar trend for the Arctic ozone hole, though the latter was not significant for every spring Arctic. In contrast, the recovery of ozone at mid-latitudes or the tropical ozone hole in the lower stratosphere was delayed by around 10 years from the declining trend of tropospheric ODSs [4, 5, 32, 33, 39]. Remarkably, the observed data also showed the formation of three 'temperature holes' corresponding to the ozone holes over the Antarctic, tropical and Arctic since the 1980s, respectively [33]. The lower-stratospheric temperature variations in the three 'temperature holes' are well reproduced by the CRE equation with the level of anthropogenic halogenated ODSs (mainly CFCs) and the CR intensity in the stratosphere as the only two variables. This is also true for the global lower-stratospheric temperature (GLST), as will be demonstrated by observed data presented in this paper.

It is generally agreed that the change in global stratospheric temperature should mirror the change in GMST. Given the observations outlined above, it is crucial to understand why there was a warming pause in 2000–2015 and why the GMST has risen again in recent years. This paper aims to address these critical questions using substantial observation datasets, including GLST, troposphere and stratosphere temperature climatology, land surface air temperature, GMST, sea ice extent (SIE) and snow cover extent (SCE), as well as physical model calculations of GLST and GMST.

## 2. Data and Methods

We choose to use observed data as original as possible when available instead of 'adjusted' or 'processed' data based on understanding in current climate models. The following datasets are used for the present study. As used in our recent study [33], the same multiple datasets obtained from ground- and satellite-based measurements of time-series annual mean lower-stratospheric temperature anomalies in the lower stratospheres (100–30 hPa) since the 1960s, including NOAA's MSU UAH [40] and RSS [41] satellite datasets (Channel 4), EUMETSAT's ROM SAF's radio occultation (RO) satellite datasets [42] and NOAA's radiosonde-based Ratpac-B time-series dataset [43], are used to show the time-series lower-stratospheric temperatures or the temperature climatology in the troposphere and stratosphere of the global (90° S–90° N). A representative zonal mean latitude–altitude distribution of $CF_2Cl_2$ (CFC-12) is obtained from the NASA UARS (CLEAS) dataset. For global land surface air temperature and GMST datasets, we use those obtained from the UK Met Office (CRUTEM4.6, CRUTEM5.0 and HadCRUT4.6) [44,45], while land surface air temperatures of regions or individual countries are obtained from Berkeley Earth, UK Met Office (HadCET) or NOAA (ClimDiv CONUS). The sea ice extent (SIE), snow cover extent (SCE) and ENSO datasets are obtained from NOAA. Cosmic ray data are obtained from ref. [32]. Updated concentrations of $CO_2$ and halogenated GHGs as well as calculated radiative forcings of $CO_2$ given in current climate models are obtained from the 2021 IPCC AR6 Report [8].

Although the newest IPCC AR6 Report [8] has provided good summaries of recent progress in observations, it has led to changes in historical observed GMST datasets. For example, the IPCC AR5 [6] assessed estimate for historical warming between 1850–1900 and 1986–2005 was 0.61 (0.55 to 0.67) °C, while it has been changed to 0.69 (0.54 to 0.79) °C in the AR6 Report for this same warming period due to 'changes in observational understanding' [8]. In the latter, similar 'adjustments' since 2013 can be found in most GMST datasets. This could create errors, given that current climate models still have significant discrepancies from observations. Nevertheless, there are no major differences between observed data of GLST, the temperature climatology in the troposphere and stratosphere, GMST, land surface air temperature and SIE used in this paper and the IPCC AR6 Report [8], except that the latter shows declining trends in the SCE in Northern Hemisphere (NH) in April over the 1922–2018 period (see Figure 2.22 of ref. 8) and for all seasons over the 1981–2018 period (see Figure 9.23 of ref. 8), whereas this paper using the NOAA Climate Data Record provided by the Rutgers University Global Snow Laboratory [46, 47] shows that the annual mean SCE in NH or North America has stabilized since the mid-1990s. However, new insights into mechanistic understanding of these climate changes are provided in this paper.

To compare our theoretical results with observed GMST data, it is necessary to remove the natural El Niño and volcanic effects from observed GMST data. There was no significant volcanic effect during 2000–2021, while ENSO was the largest source of year-to-year variability [8]. We simply adopt the empirical model developed by Lean and Rind [21, 22], which played an important role in identifying that the warming in the late half of last century was due to anthropogenic influences. In this model, lags are 4 months for ENSO, 6 months for volcanic aerosols and 120 months (10 years) for the anthropogenic force. The latter (the 10-year lag) was originally chosen to maximize the explained variance in the Lean–Rind model, which turns out to agree excellently with the observed ozone recovery trend at tropical and mid-latitudes delayed from the tropospheric halogenated ODS peak [4, 5, 32, 33, 39]. The natural contributions to the GMSTs were 0.2 °C warming during major ENSO events in 1997–98 and about 0.3 °C cooling in 1992 following the large Pinatubo volcanic eruption [21, 22]. These values are used to remove ENSO and volcanic effects from observed GMST data without further optimization, with details given in Supplementary Materials (see Figures S1 and S2).

To calculate the contribution of halogenated GHGs to GMSTs, we use our CFC-warming quantum physical model with details given previously [5]. Note that our physical model includes no parameter but the halogenated-gas specific equilibrium climate sensitivity ($\lambda_c^{halo}$), which was determined by the product of the climate sensitivity factor ($\alpha$) of halogenated GHGs in the atmospheric window at wavelengths of 8–13 μm and the feedback amplification factor ($\beta$) [5]. Both $\alpha$ and $\beta$ were, respectively, determined from quantum physics understanding of the satellite-measured atmospheric transmittance spectrum in the atmospheric window, at which halogenated GHGs have strong infrared absorption bands and well-observed surface temperature variations during solar cycles. Through this observational determination, we obtained the value of $\lambda_c^{halo}$ (=$\alpha\beta$), which is equal to 1.77 K/(W.m$^{-2}$) [5]. With this equilibrium climate sensitivity $\lambda_c^{halo}$ for halogenated GHGs, given radiative efficiencies and a lag of 10 years of halogenated GHGs, our physical model calculations using sole inputs of measured and projected atmospheric concentrations of halogenated GHGs directly give time-series contributions of the latter to GMST.

## 3. Results and Discussion

### *3.1. Global Lower-stratospheric temperature (GLST)*

Figure 1a shows time-series lower-stratospheric temperatures of the global (90° S–90° N) from multiple data sources obtained from ground- and satellite-based measurements [40-43]. The results show that although there are large discrepancies between ground- and satellite-based measured datasets prior to 1995, the lower-stratospheric temperatures from all datasets consistently exhibit a clear drop between the 1970s and 1995 and have become constant since the mid-1990s, with no significant change over the past 27 years. This is consistent with the previous observations [7, 28, 29, 32, 33] and those reported in the newest IPCC AR6 [8]. Namely, the AR6 states: "It is virtually certain that the lower stratosphere has cooled since the mid-20th century. However, most datasets show that lower-stratospheric temperatures have stabilized since the mid-1990s with no significant change over the last 20 years" (page 2–48, chap. 2).

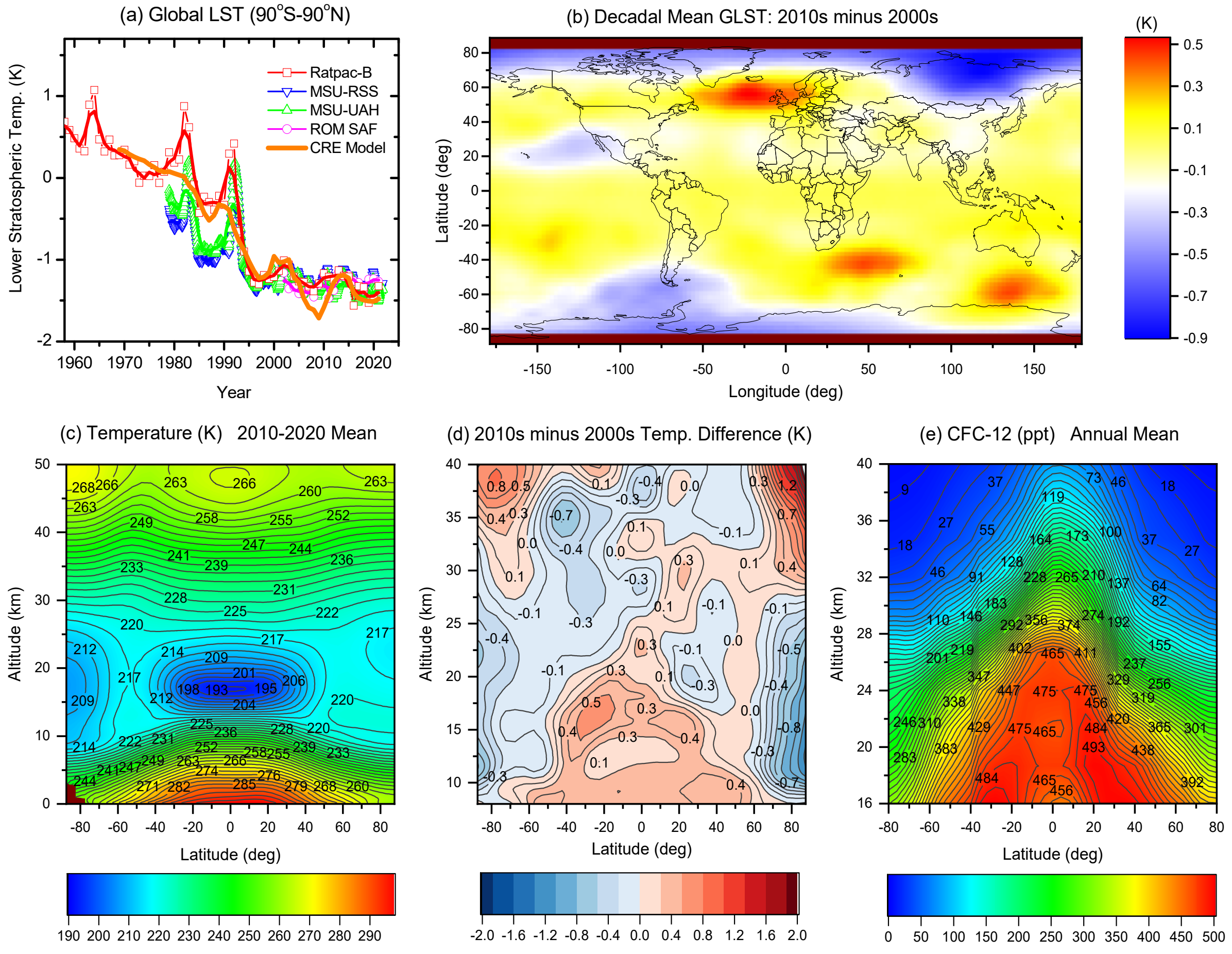

**Figure 1.** (**a**): Time-series annual mean lower-stratospheric temperature anomaly datasets of the global (90° S–90° N) since 1958, obtained from multiple ground- and satellite-based data measurements (Ratpac, MSU-UAH, MSU-RSS and ROM SAF), as well as temperatures calculated by the CRE equation (thick solid lines in orange, see text). Further, also shown are the 3-year smoothing (thick solid lines in colors) to observed temperature anomalies (symbols). Note that for comparisons, the lower-stratospheric temperature anomalies are offset to the same level in the 2000s–2010s, but this offsetting has no effects on their long-term trends. (**b**): Map for the decadal mean global lower-stratospheric temperature (GLST) difference of the 2010s (2010–2020) minus the 2000s (2000–2010). (**c**): Decadal mean zonal mean latitude–altitude distribution of the temperature climatology in the troposphere and stratosphere in the 2010s (averaged from 2010 to 2020). (**d**): Difference in decadal mean zonal mean latitude–altitude distribution of the temperature climatology at altitudes of 8–40

km of the 2010s (2010–2020) minus the 2000s (2002–2010). (**e**): Representative zonal mean latitude–altitude distribution of the $CF_2Cl_2$ (CFC-12) concentration obtained from the NASA UARS's CLEAS dataset.

Given transport lags of 1 and 10 years for ODSs transported from the troposphere/surface to the lower stratosphere over the polar regions and the near global (60° S–60° N) or the tropics, respectively, according to the CRE mechanism and observations [4, 5, 29, 32, 33, 39], we obtain a transport lag of approximately 9 years for ODSs in the global mean lower stratosphere to model the GLST using the CRE equation. For simplicity, we use the same CRE reaction constant $k_{Tro}$ to calculate the GLST as that used for calculating the lower-stratospheric temperature over the tropics [33], except that here, we use a different transport lag of 9 years for ODSs. That is, no optimizations to obtain the best fitting results are performed. This should be a good approximation for calculating the GLST, considering that the total area of the polar regions only takes about 13% of the globe and the tropical ozone/temperature hole is all season in contrast to the polar holes that appear in springtime only. The calculated results of the GLST are also shown in Figure 1a, which match the observed data well, as we expected. Overall, the calculated GLST curve shows excellent agreement with the 3-year averaged observed curve. Despite the obvious simplifications and approximations in the above CRE-model calculations, the results in Figure 1a clearly demonstrate that the GLST has well been controlled only by the level of halogenated ODSs and the CR intensity and that lower-stratospheric cooling has ceased since the mid-1990s, with no significant change over the past 27 years. This is consistent with the measured trend in ODSs and the observed recoveries in ozone depletion shown previously [4, 5, 13, 29, 32, 33, 37, 38]. However, the results in Figure 1a cannot explain why the GMST has risen again in recent years (since 2015).

To find an answer to the above puzzling question, it is important to obtain more detailed information from the map of GLST, which is made from the available MSU-UAH satellite datasets [40]. The map for the decadal mean GLST difference of the 2010s minus the 2000s is shown in Figure 1b. It is now clearly revealed that the cooling in the lower stratosphere over the tropics and mid-latitudes has started to become warming over the past decade, which is consistent with the observed recovery of the tropical $O_3$ hole [33]. However, lower-stratospheric cooling remains significant in some local high-latitude areas of the Arctic and Antarctic. The enhanced stratospheric cooling is most marked in North and Northeast Russia, extending to the Far-East region of Russia and Alaska in the USA; it also occurs at some areas in Antarctica, especially at its west side. It is obvious that these lower-stratospheric coolings at the polar regions do not agree with the decreasing trends of ODSs and associated ozone depletion [4, 5, 32, 37, 38]. It must originate from a different mechanism, which will be revealed further by observed data below.

### *3.2. Temperature Climatology in the Troposphere and Stratosphere*

To resolve the above mystery, we further investigate the zonal mean latitude–altitude distribution of the temperature climatology in the troposphere and stratosphere for the period of 2002–2020, obtained from the high-quality ROM SAF RO satellite datasets [42]. Similar to the results for the 2000s (2002–2010), reported previously [33], the decadal mean zonal mean latitude–altitude distribution of the temperature climatology for the 2010s (2010–2020) is now plotted in Figure 1c, which shows clearly the three 'temperature holes' corresponding to the ozone holes over the Antarctic, tropics and Arctic, respectively. The global tropospheric–stratospheric temperature differences at altitudes of 8–40 km of the 2010s minus the 2000s are shown in Figure 1d, in which very interesting results are now found. First, the overall temperature difference (warming) pattern in the troposphere and stratosphere remarkably resembles the representative atmospheric distribution pattern of CFCs shown in Figure 1e. Second, consistent with the MSU satellite data shown in Figure 1b, lower-stratospheric cooling in the tropics and mid-latitudes has been reversed to become warming over the past decade. Third, this reversal has not yet occurred over the Antarctic and there has even been a small increase in lower-stratospheric cooling at latitudes of 70°–90° S, in contrast to the observed recovery of the Antarctic $O_3$ hole [4, 5, 32, 37, 38]; lower stratospheric cooling over the Arctic is significantly enhanced at latitudes of 60°–90° N, in spite of no enlarging in the Arctic $O_3$ hole. Fourth, most remarkably, upper-stratospheric *warming* has been observed to be significant at both the southern hemisphere (SH) and NH high latitudes and it has been more significant (up to a 2.8 K increase) at NH than SH high latitudes over the past decade. These observed results of altitude profiles of the temperature differences are more clearly shown in Figure 2a,b, which obviously contradict the upper-stratospheric cooling predicted by the $CO_2$-dominant warming climate models with the observed increasing annual growth rate of atmospheric $CO_2$ but are in excellent agreement with the CFC-dominant warming model [4, 5, 14]. According to the latter model of global surface warming and the CRE mechanism of stratospheric ozone depletion, upper-stratospheric cooling (surface warming) should have been reversed first at high latitudes with decreased levels of atmospheric CFCs, as CFCs are more effectively destroyed by the stronger CRE reaction at higher latitudes [4, 5, 32, 35] (see also Figure 1e). Note also that this reversal of upper-stratospheric cooling is counteracted by the recovery of $O_3$ depletion at high latitudes, as ozone itself is an effective GHG. Such an effect from $O_3$ recovery should be much smaller at NH than SH high latitudes because $O_3$ loss over the Arctic has been far less than over the Antarctic since the late 1970s. Therefore, upper-stratospheric warming should be more significant at NH than SH high latitudes. These major features of the warming model of halogenated GHGs (CFCs) are exactly consistent with the observed data shown in Figures 1d and 2a,b. However, the observed data in Figures 1b,d and 2a,b also consistently indicate that there is an active mechanism that is affecting the local/regional lower-stratospheric temperatures at some polar regions (60°–90° N and 70°–90° S), rather than the warming mechanism of halogenated GHGs.

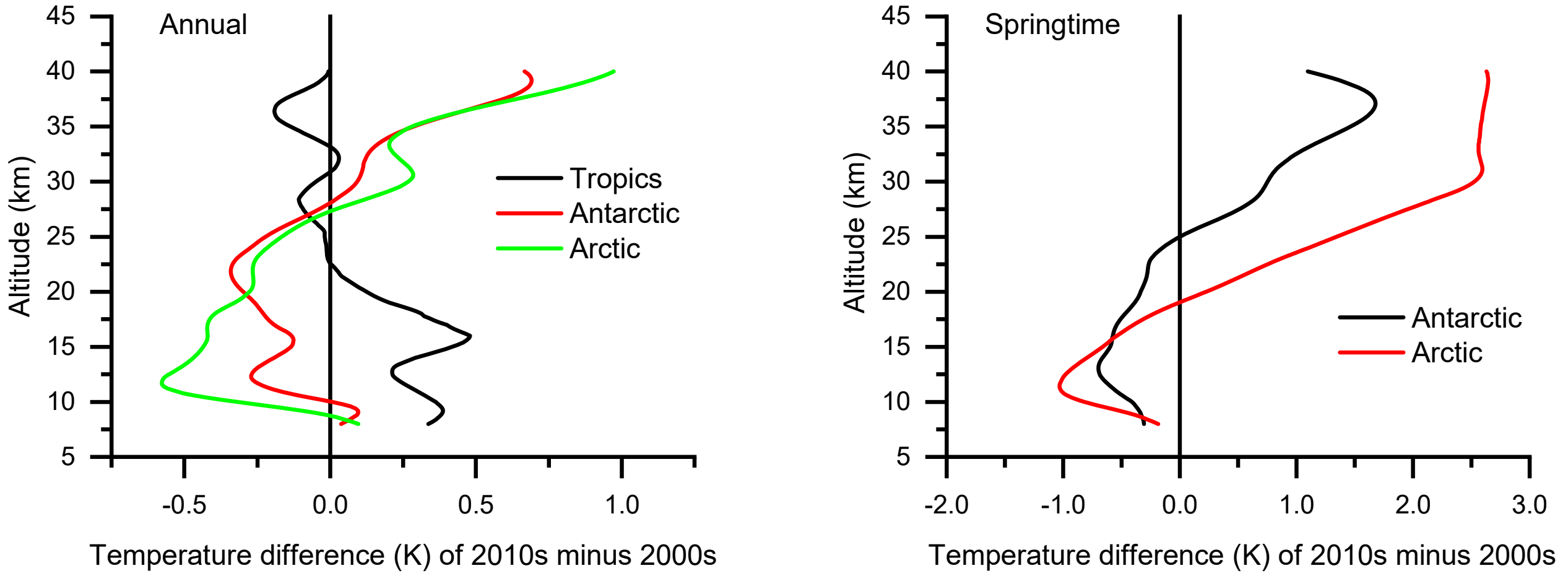

**Figure 2.** (**a**,**b**): Altitude profiles of the temperature differences at altitudes of 8–40 km of the 2010s (2010–2020) minus the 2000s (2002–2010) for the annual tropics (30° S–30° N), annual and springtime Antarctic (60°–90° S) and Arctic (60°–90° N).

### *3.3. Global Land Surface Air Temperatures*

Since the stratospheric temperature change mirrors the surface temperature change, we turn to investigate the detailed changes in maps of global land surface air temperature. Here, we focus on analyzing changes in global land surface air temperature instead of GMST for the following two reasons: (i) the temperature rise over land has been larger than over the oceans since 1850; (ii) GMST is a combination of land surface air temperature and sea surface temperatures (SSTs), whereas there are distinct differences between SSTs and marine air temperatures (MATs), leading to long-term trend differences between GMST and global surface air temperature that is the combination of land surface air temperature and MATs by ±10% with poor theoretical understanding [8]. Therefore, we focus on studying the maps of land surface air temperatures, which are made from the UK Met Office's CRUTEM5 global gridded monthly (land) air temperature dataset [45]. The maps for the land surface air temperature differences of 2000–2020 minus 1950–1975 and 2010–2020 minus 2000–2010 are shown in Figure 3a,b, respectively. Figure 3a clearly shows a significant and highly inhomogeneous global warming in the late half of last century, as widely reported in the literature [8]. In contrast, Figure 3b shows that, consistent with the map of GLST (Figure 1b), significant surface warming has only continued in north and northeast coastal regions of Russia and in Alaska in the USA over the past decade, while surface warming has been ceasing or slightly reversing in most areas of the globe. Unlike the GLST result in Figure 1b showing lower-stratospheric cooling at the west side of Antarctica, the continued surface warming in west Antarctica is not visible in Figure 3b, which is likely due to very few measurement stations in the region.

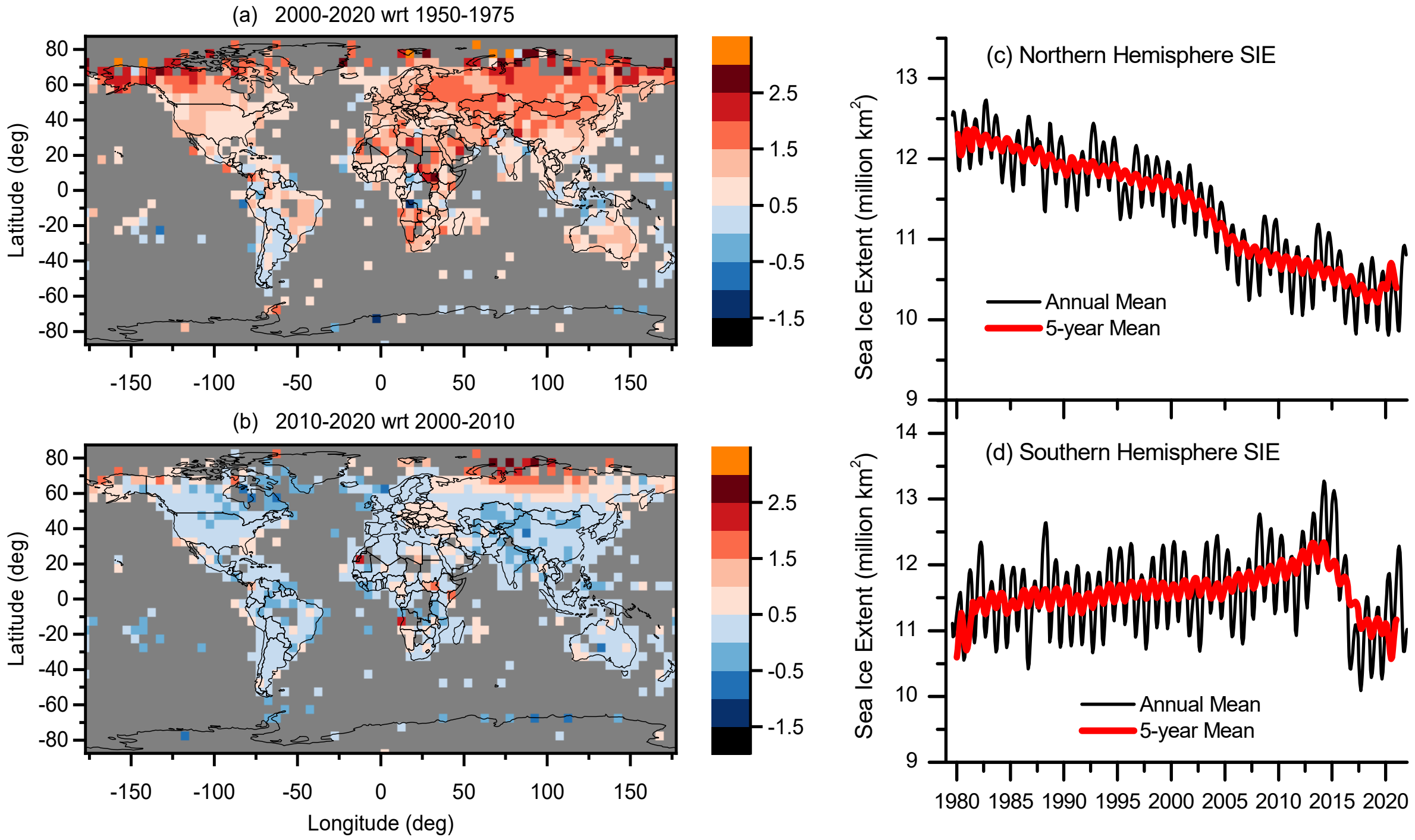

**Figure 3.** (**a**,**b**): Maps for global land surface air temperature differences of 2000–2020 minus 1950–1975 and 2010–2020 minus 2000–2010. (**c**,**d**): Time series sea ice extent (SIE) in the Southern Hemisphere (SH) and Northern Hemisphere (NH) during 1979–2022.

### *3.4. Evidence of the 'Arctic Amplification (AA)' Mechanism for Polar Warmings*

To reveal the mechanism for the regional continued warmings at the polar coasts, it is worth noting the cause of the so-called 'Arctic amplification (AA)' phenomenon, which refers to the observed much faster warming in the Arctic than the rest of the world over the last half century. Although the precise mechanism for AA is still under debate [48], there is an interesting mechanism proposed by Dai et al. [49]: AA is closely related to the surface albedo feedback associated with sea-ice loss, leading to increased outgoing longwave (LW) radiation and heat fluxes from newly opened waters. Their simulations showed that AA occurs primarily in the cold season (from October to April) due to the extra LW radiation and sensible and latent heat release from the newly opened waters, which are 10–30 °C warmer than sea-ice surfaces in Arctic winter and only over areas with significant sea-ice loss and to largely disappear when the sea ice is fixed or melts away under GHG-induced warming. Here, NH and SH sea-ice extent (SIE) data are shown in Figure 3c,d, which confirm the continued sea-ice loss over the Arctic and the sudden melting at Antarctica since around 2015. Over the past two decades, the Arctic surface temperature has been 2–3.5 °C higher than that in 1950–1975 (Figure 3a). Thus, even a small reversal (decrease) in GMST will not cause an immediate stopping in ice melting in the Arctic. By contrast, the SIE at Antarctica had a gradual *increase* from 1979 to 2015 and had an abrupt drop around 2015. This difference in SIE change between the Arctic and Antarctic (Figure 3c,d) can be well explained by the well-known much larger $O_3$ loss over the Antarctic, which counteracted the warming effect caused by man-made GHGs. Another factor is that the CRE destruction of halogenated ODSs is more effective over the SH than the NH, especially at high latitudes (see Figure 1e). Thus, the observed surface warming was much milder over the SH than the NH, as seen in Figure 3a. The current recovery of the Antarctic $O_3$ layer is changing this trend. Note that the small peak in the SH SIE in 2020–2021 in Figure 3d is consistent with the observed Antarctic $O_3$ hole maxima, precisely corresponding to the CR peak in the recent solar cycle [32]. The AA mechanism associated with sea-ice loss agrees well with the observed results shown in Figures 1b,d, 2a,b and 3a–d. Even more obviously, it agrees with the observed seasonal NH land surface air temperature differences of 2010–2020 minus 2000–2010, which are shown in Figure 4a–d. The latter shows that continued regional warmings at the Arctic coasts, indeed, only occur in the seasons of DJF (December, January and February) and MAM (March, April and May) but not JJA (June, July and August) and SON (September, October and November), in good accord with the expectation from the sea-ice-loss-caused AA mechanism [49].

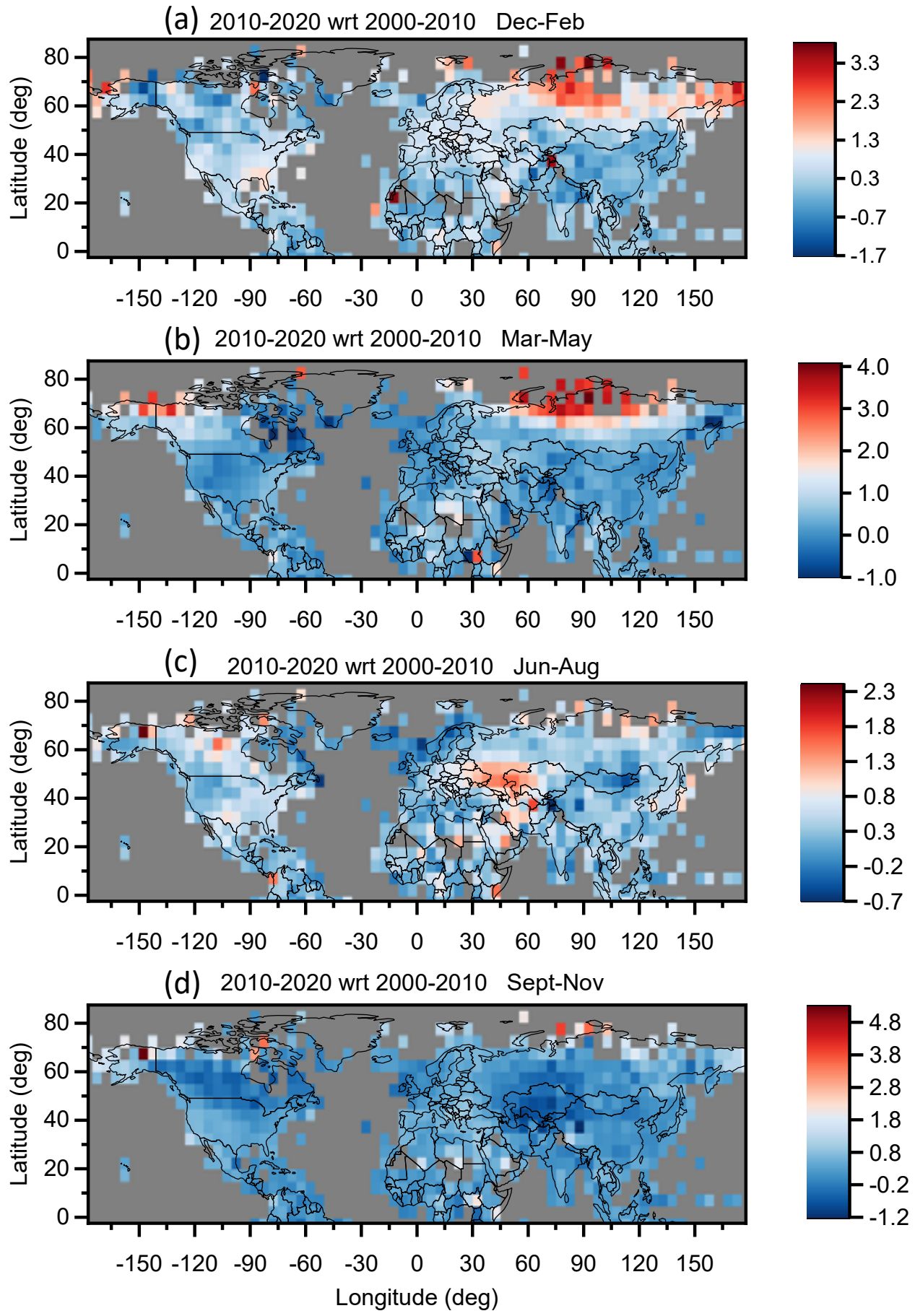

**Figure 4.** (**a**–**d**): Maps for seasonal NH land surface air temperature differences of 2010–2020 minus 2000–2010 in DJF, MAM, JJA and SON.

To provide further evidence of the above AA mechanism of sea-ice loss for continued Arctic warming on a microscopic geographical scale, we show time-series annual-mean surface temperatures of three representative coastal states or observation stations, in which sea-ice melting occurs versus three representative inland states with no sea-ice melting in Russia and time-series annual, 5-year and 10-year mean surface temperatures by area-weighted averaging from the 12 coastal states or stations (Arkhangel'sk, Chukot, Kamchatka, Khabarovsk, Koryak, Krasnodar, Maga Buryatdan, Murmansk, Nenets, Sakha, Taymyr and Yamal-Nenets) versus the 39 inland states or stations (Aga Buryat, Altay, Bashkortostan, Buryat, Chelyabinsk, Chita, Chuvash, Gorno-Altay, Irkutsk, Kemerovo, Khakass, Khanty-Mansiy, Kirov, Komi-Permyak, Kostroma, Krasnoyarsk, Kurgan, Mariyel, Mordovia, Nizhegorod, Novosibirsk, Omsk, Orenburg, Penza, Perm', Primor'ye, Sakhalin, Samara, Saratov, Sverdlovsk, Tambov, Tatarstan, Tomsk, Tuva, Tyumen', Udmurt, Ul'yanovsk, Ust-Orda Buryat and Yevrey) in Russia in Figure 5. In the lower panel in Figure 5, linear fits to the observed temperature data after 2000 are performed and the slopes ΔT (changed temperature per year) and the $R^2$ values (the coefficient of determination, COD) are given. Although the AA warming of sea-ice loss at the coastal regions is expected to influence the surface temperatures at the inland regions in a country to some extent, the observed data show clearly a drastic difference: over the past two decades, the surface temperatures at sea-ice melting states exhibit sharp rises with ΔT = 0.53 ± 0.02 °C/year and $R^2 = 0.67$, whereas the surface temperatures at inland states have been fairly flat or much slower increasing with ΔT = 0.18 ± 0.03 °C/year and $R^2 = 0.13$. For the latter, the very low $R^2$ value close to zero indicates no significant trend over the past two decades. From the observed results shown in Figures 1–5, we can make a solid conclusion that regional warmings at the Arctic coasts of Russia and Alaska (USA) and some areas in Antarctica are clearly due to continued/new sea-ice loss.

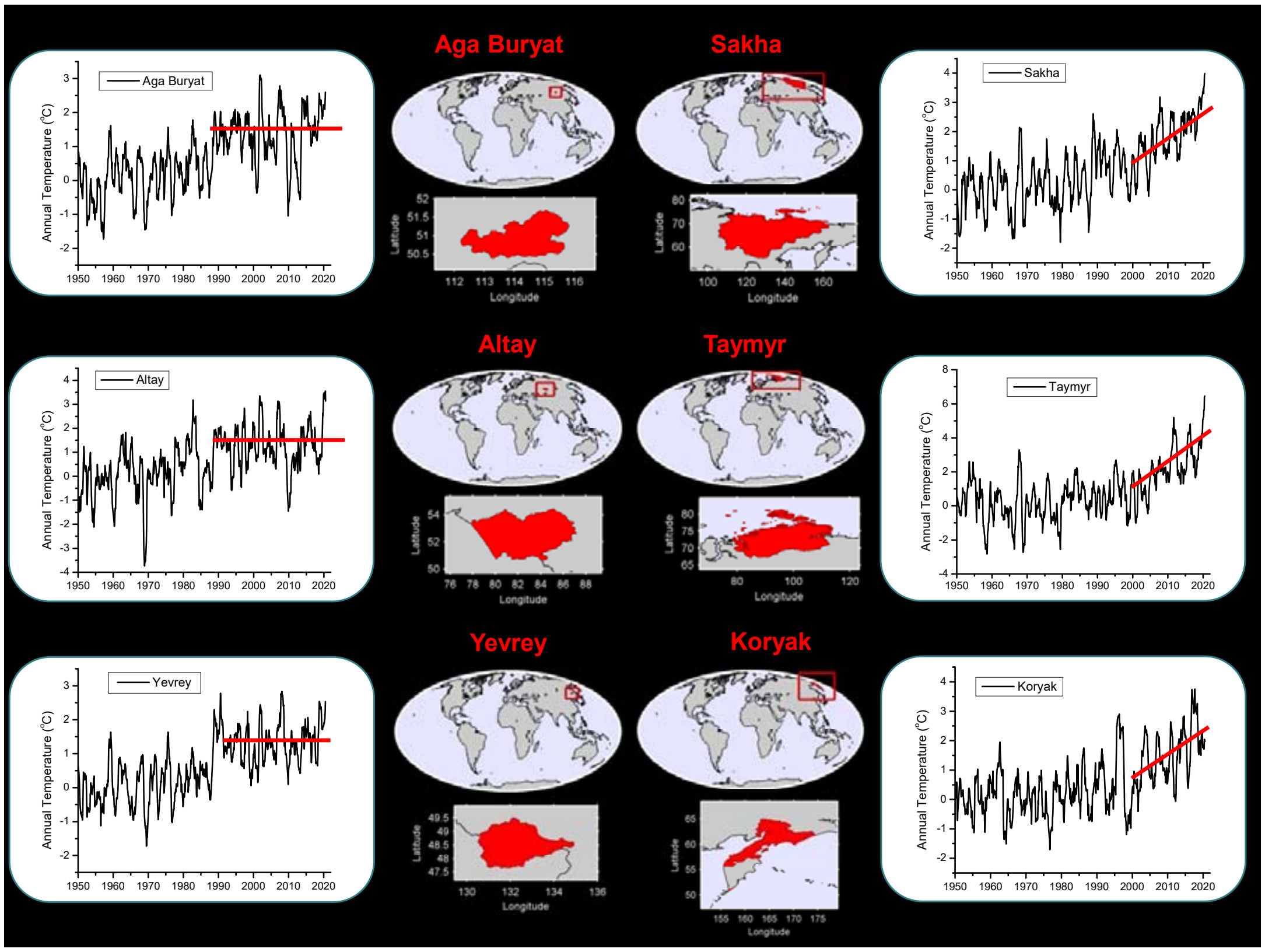

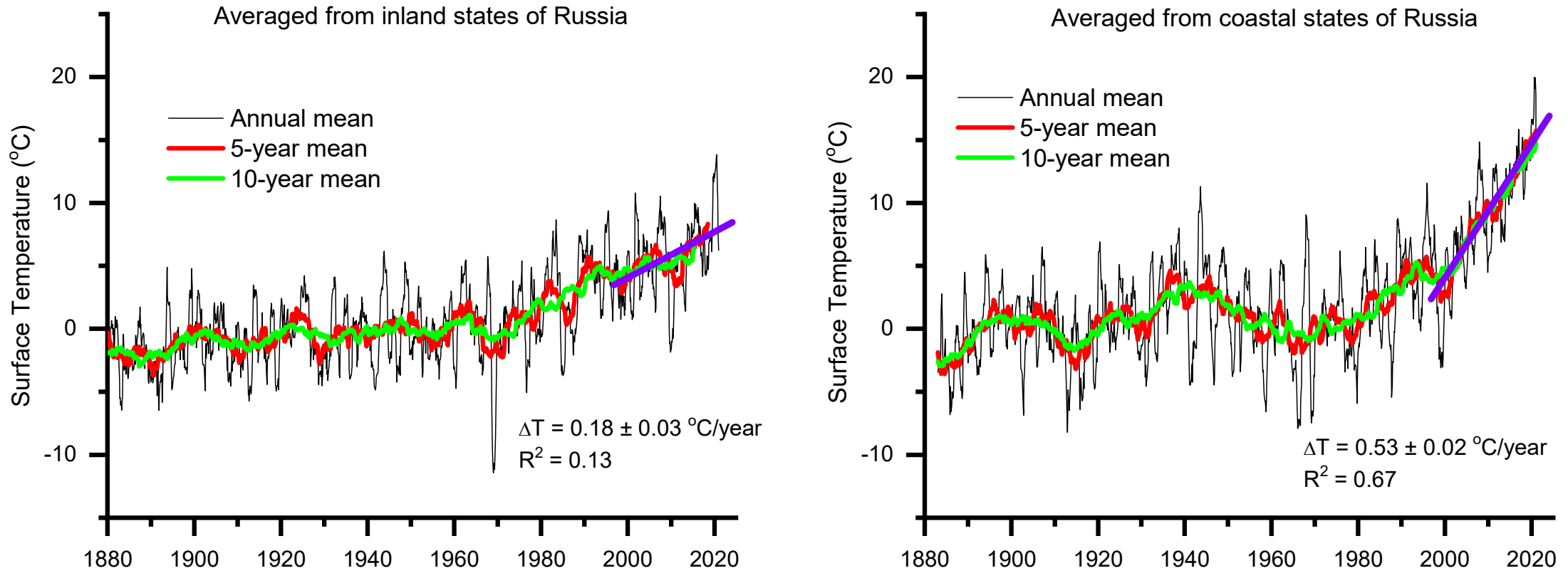

**Figure 5.** (Upper panel) Time-series annual surface temperatures of three representative inland states (Aga Buryat, Altay and Yevrey) that have no sea-ice melting versus three representative coastal states or observation stations (Sakha, Taymyr and Koryak) in which sea-ice melting occurs in Russia (the red lines roughly showing the change trends). (Lower panel) Time-series annual, 5-year and 10-year mean surface temperatures averaged from the 39 inland states or stations with no sea-ice melting versus the 12 coastal states or stations in which sea-ice melting occurs in Russia (see the text), where linear fits to the observed data after 2000 are given (the violet lines), with the slopes ΔTs (changed temperatures per year) and the $R^2$ (COD) values indicated.

### *3.5. Global Warming Pause at NH Extratropic Excluding Russia and Alaska*

All the above observed data point to the likely mechanism that human-made halogen-containing GHGs (CFCs, HCFCs, HFCs, and PFCs) have played a major role in global warming. For this mechanism, the main contribution arose from CFCs (the major ODSs) up to the 2000s, while the contributions of HCFCs, HFCs and PFCs (non-ODSs but GHGs) are increasing [5, 7, 8]. If this mechanism is dominant, then the stopping or reversal of global warming should first occur at high-latitude regions if there were no sea-ice-loss-caused AA effect, as discussed above. To confirm this, time-series land surface air temperature at NH extratropic (latitudes 30° N–90° N), excluding Russia and Alaska since 1950, obtained from the UK Met Office's CRUTEM4.6 dataset, is shown in Figure 6a, in which the natural El Niño and volcanic effects are removed, with details given in Supplementary Materials (Figure S2). The result in Figure 6a seems to indicate a slowing down or cessation in surface warming; that is, the surface temperature has reached a plateau since around 2005. This is qualitatively consistent with the total effective radiative forcing (ERF) trend of halogenated GHGs [5, 8], as shown in Figure 6b. This is in sharp contrast to the rising trend for the calculated $CO_2$ ERFs given in current climate models [8], due to the increased annual growth rate of the measured atmospheric $CO_2$ level, as shown in Figure 6c. This slowing down or stopping of surface warming is also validated by the following observations: (i) The surface temperature changes in North America (Canada, contiguous USA and Greenland) show a similar warming-stopping behavior, as shown in Figure 6d. (ii) Consistently time-series snow cover extent (SCE) data over NH and North America since 1967 plotted in Figure 6e,f, obtained from the NOAA Climate Data Record [46, 47], show consistently that SCEs have clearly stabilized since ~1995. (iii) A quite similar result with no temperature increase over the past two decades can also be found from the UK Met Office's central England temperature (CET) dataset for the period of 1659–2021, which is the longest instrumental record of temperature in the world [50], as shown in Figure 7. (iv) Surface temperature changes in North Europe (Sweden, Norway, Finland, UK, Ireland and Iceland) and North Asia, including 11 countries plus 12 north and west provinces of China, show similar warming slowing or stopping phenomena, as shown in Figure S3. Note that for these regional land surface air temperature changes, only the original observed data are plotted, that is, the natural El Niño and volcanic effects are *not* removed as it is hard to do so for local regions or individual countries.

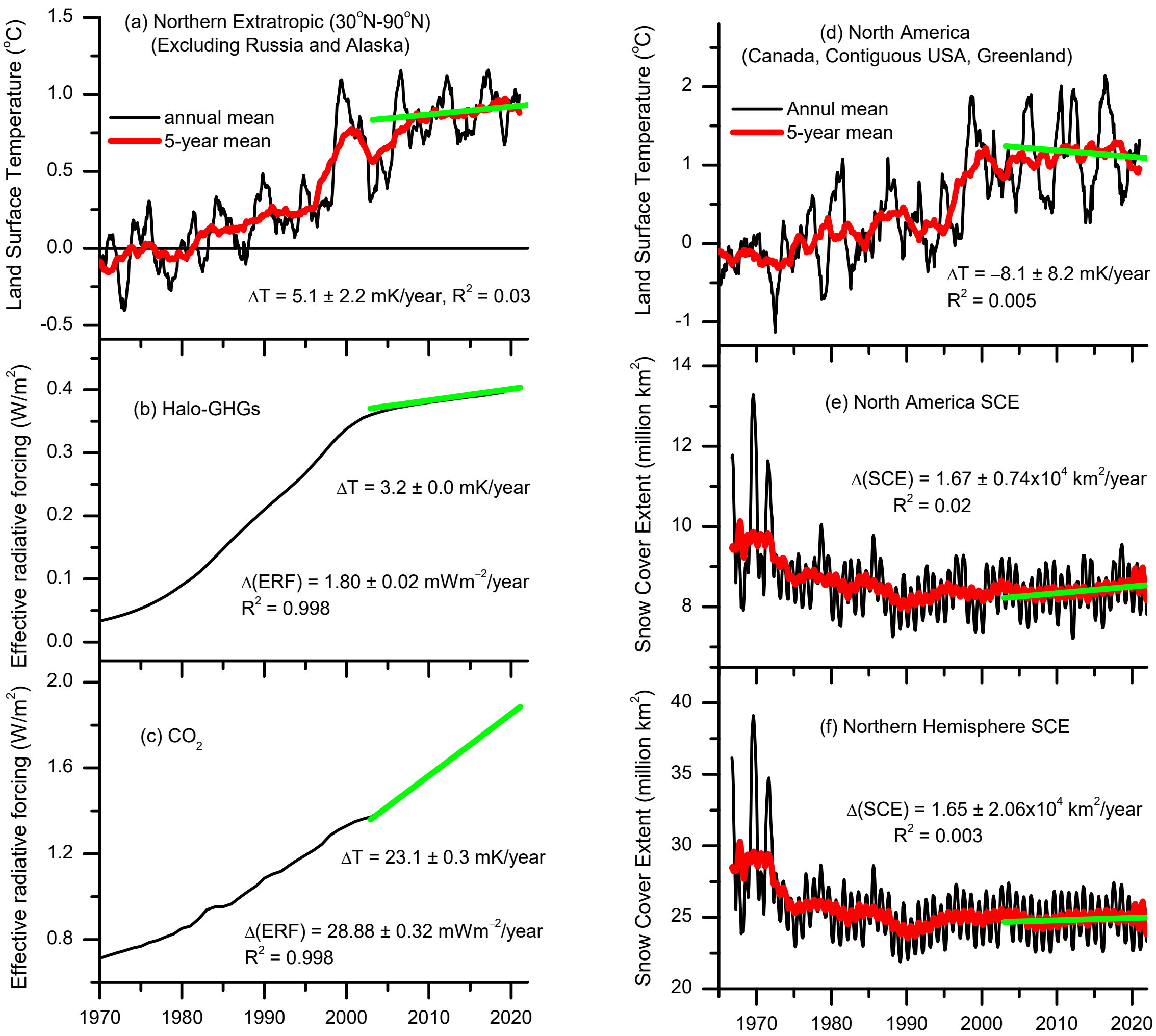

**Figure 6.** (**a**): Time-series measured land surface air temperature at Northern Hemisphere (NH) extratropic (30° N–90° N) excluding Russia and Alaska. (**b**,**c**): Time-series calculated effective radiative forcings (ERFs) of atmospheric $CO_2$ and halogenated greenhouse gases (halo-GHGs), given in climate models [5, 8]. (**d**): Time-series measured surface temperatures in North America (Canada, contiguous USA and Greenland). (**e**,**f**): Time-series snow cover extent (SCE) in North America and NH during 1968–2021. For each dataset in (**a**–**f**), a linear fit to the observed data after 2005 is given (the green line), with the slope (changed amount per year) and the $R^2$ (COD) value indicated.

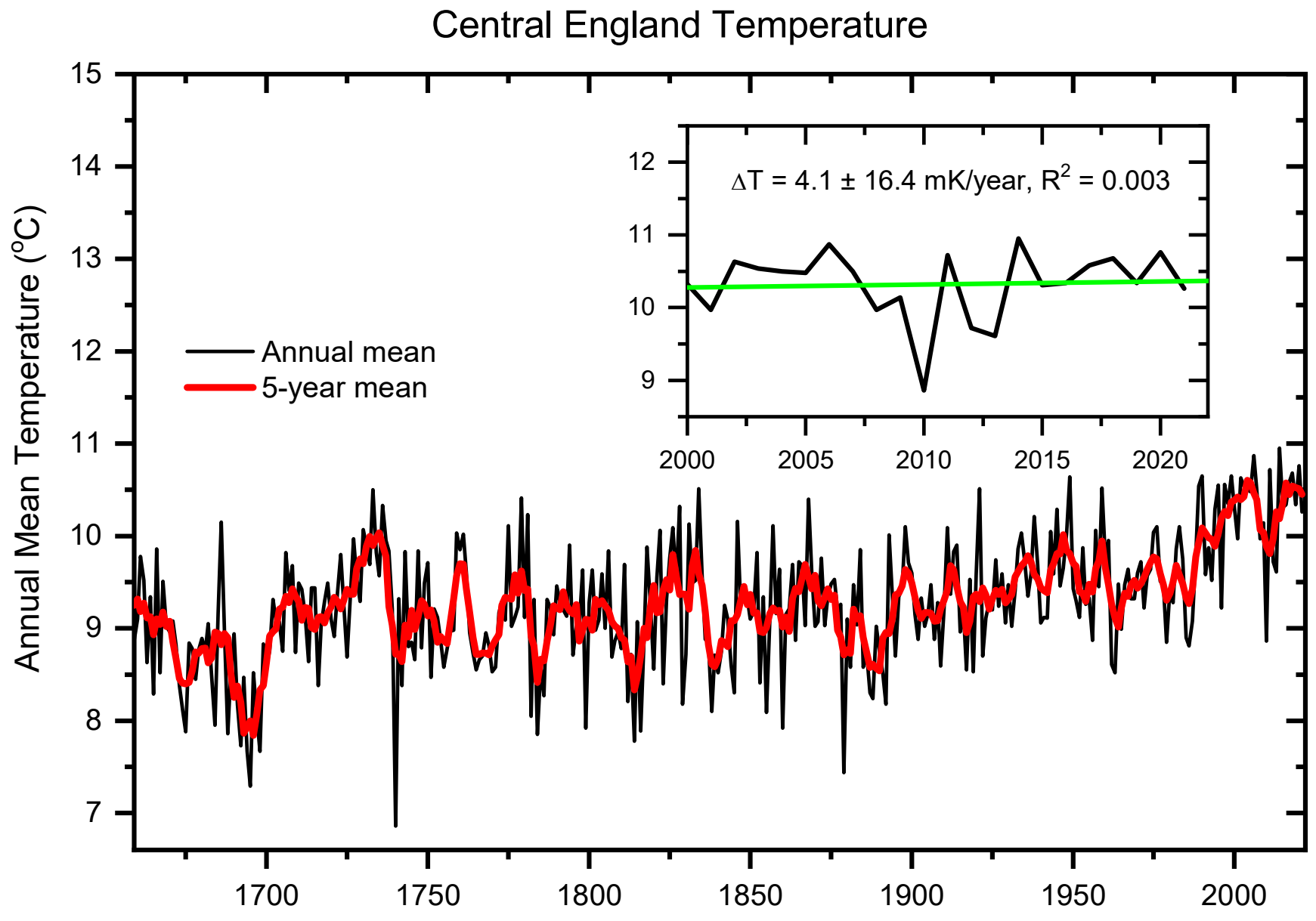

**Figure 7.** Annual and 5-year mean temperature of the central England representative of a roughly triangular area of the United Kingdom enclosed by Lancashire, London and Bristol for the period of 1659-2021, which is the longest instrumental record of temperature in the world. The inset shows annual temperatures in 2000-2021 (the black curve) and a linear fit to the observed data (the green line), with the slope ΔT (changed temperature per year) and $R^2$ value indicated.

Furthermore, linear fits to the observed data after 2005 are performed in Figures 6a–f and 7, in which the slope (changed amount per year) and the $R^2$ value for each fit are given. The fitted results show that the surface temperature at NH extratropic is nearly flat with ΔT = 5.1 ± 2.2 mK/year and $R^2 = 0.03$ (Figure 6a). Interestingly, the fit to the data of ERFs arising from atmospheric halogenated GHGs, calculated from all climate models [5, 8], gives a slope Δ(ERF) = 1.80 ± 0.02 $mWm^{-2}$/year and $R^2 = 0.998$ (≈ 1.0) in Figure 6b. With the given equilibrium climate sensitivity $\lambda_c^{halo}$ specifically for halogenated GHGs in Section 2, obtained from our warming model [5], one can straightforwardly find the corresponding temperature-rising rate to be 3.2 ±0.0 mK/year for the period of 2005–2020, which is in good agreement with the observed value shown in Figure 6a. In contrast, the fit to the calculated ERFs for $CO_2$ given from the IPCC-accepted climate models [8] leads to a slope Δ(ERF) = 28.88 ± 0.32 $mWm^{-2}$/year, one order of magnitude larger than that of all halogenated GHGs in total. In the current IPCC's AR6, climate models have given the best estimate for global mean surface air temperature increase to be 3 °C, with a likely range of 2.5–4 °C by a doubling of atmospheric $CO_2$ [8]. Thus, the obtained slope is equivalent to the best estimated temperature-rising rate of 23.1 ± 0.3 mK/year, as also shown in Figure 6c. This result drastically differs from the observed temperature-rising rate in Figure 6a.

Moreover, the fitted results in Figure 6d–f also show that, consistent with the declining trend in North America's surface temperature with ΔT = –8.1 ± 8.2 mK/year (Figure 6d), both North America and NH SCEs have exhibited similarly positive (increasing) trends with Δ(SCE) = 1.67 ± 0.74 × $10^4$ $km^2$/year and 1.65 ± 2.06 × $10^4$ $km^2$/year, respectively (Figure 6e,f). A similar fitted result is also obtained for the temperature in central England with ΔT = 4.1 ± 16.4 mK/year (the inset in Figure 7). It should be noted that except for fits to the ERF data, all fits to the observed surface temperature or SCE data after 2005 give an almost zero $R^2$ value (≤0.03), indicating no significant change trend in each of these measured variables since 2005.

All the above results point to the important facts that an overturn in global warming would have started in around 2005 if there were no sea-ice-loss-caused warming AA at the polar regions and that the surface temperature and snow cover extent have been dominantly controlled by the change in the total radiative forcing of halogenated GHGs instead of $CO_2$.

### *3.6. Parameter-Free Theoretical Calculations of GMST*

To compare with the results calculated from our warming model of halogenated GHGs, we must first remove the natural El Niño and volcanic effects from the observed GMST data. For this purpose, we use the empirical model developed by Lean and Rind [21, 22], with details described in the Methods and Data section and given in Supplementary Materials (Figure S1).

In view of the observed results shown in Figures 1–7 and S3 and the previous observations by the author [4, 5, 13, 14, 32, 33] and others [9, 10, 12, 16], we use our warming model of halogenated GHGs only as the first-order approximation to calculate GMSTs, with details given previously [5]. As described in the Methods and Data section, this quantum-physics model needs no parameter, as it only needs inputs of updated atmospheric concentrations of halogenated GHGs, which are available from the new IPCC AR6 [8]. Our calculated results of GMST are shown in Figure 8, which exhibit excellent agreement with the observed GMST data since the 1950s. Interestingly, the calculated results also match nearly perfectly with the observed GMST data since 2015 if the regions in Russia and Alaska are excluded. These observed and theoretical results strongly indicated that climate changes in both global stratosphere and global surface have been primarily controlled by atmospheric ODSs and halogenated GHGs, respectively, with interesting complexities caused by associated stratospheric ozone depletion and polar sea-ice loss. Consistently, the postulated significant rise in GMST by $CO_2$ has been questioned in previous studies by others [9, 10, 12, 16] and the author [4, 5, 13, 14], and the observed data presented in Figures 1–8 and S3 do not agree with the $CO_2$-warming climate models. Particularly, the annual mean growth rate of atmospheric $CO_2$ concentration has increased from about 1.9 ppm/year in the 2000s to ~2.4 ppm/year in the 2010s and currently [8]. In striking contrast, the $CO_2$-model predictions of both upper-stratospheric cooling and accelerated surface warming on the ground are absent from the observed results. On the contrary, upper-stratospheric warming and real surface warming cessation have been observed, in excellent accord with the warming model of halogenated GHGs.

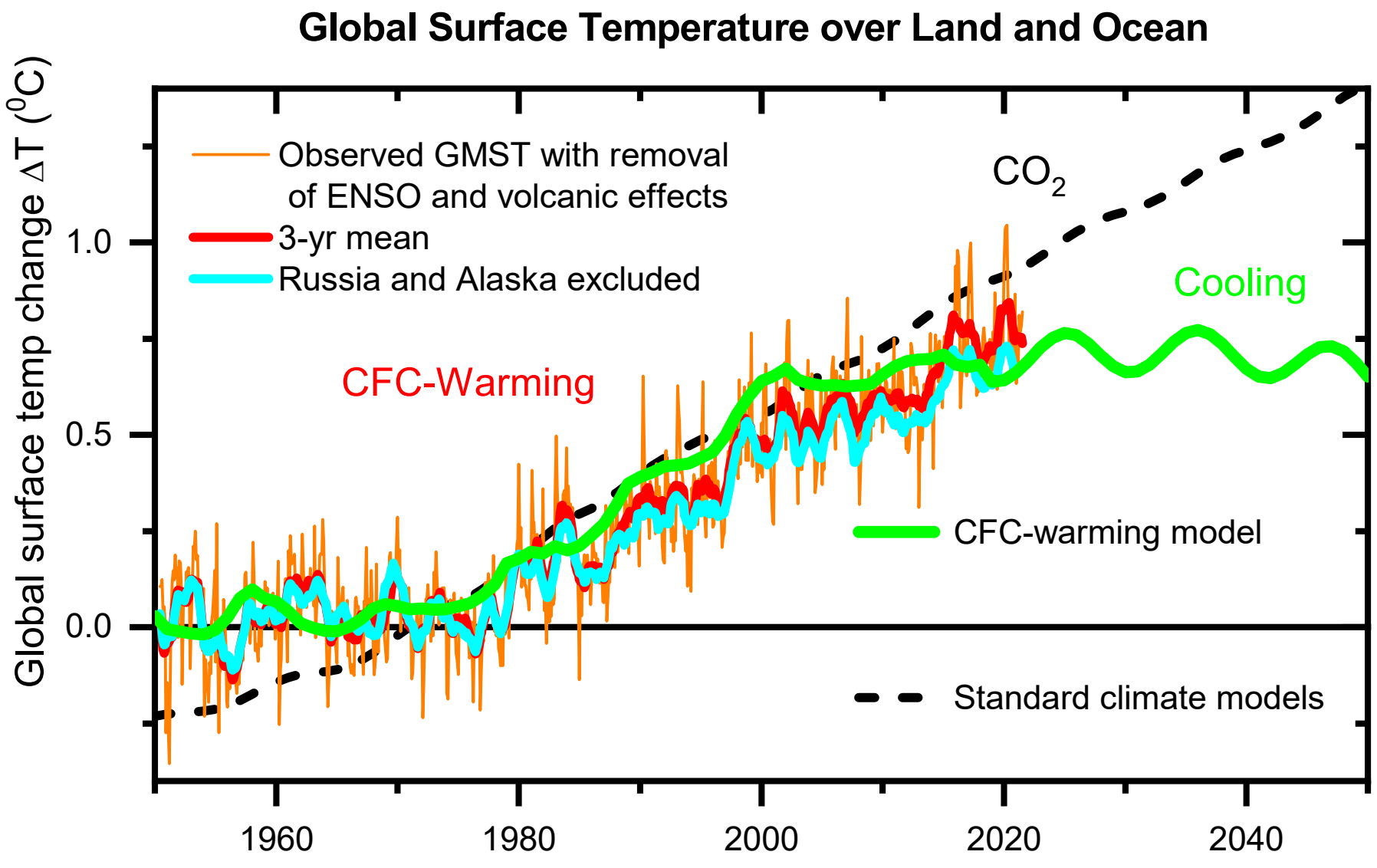

**Figure 8.** Observed and theoretical GMSTs. Observed GMST data were from the UK Met Office's combined land surface air temperature and sea surface temperature anomalies (HatCRUT4.6); the curves in red and cyan are, respectively, 3-point averages of observed data with and without the inclusion of Russia and Alaska. The theoretical GMSTs (the curve in green) were calculated by the CFC-warming model, including the contribution of all halogenated GHGs and the 11-year cyclic variation in ±0.05 °C due to solar cycles. For comparison, the simulated GMSTs by standard climate models (dashed curve in black) are also shown [5].

## 4. Conclusions and Implications

This study provides new insights in our understanding of global climate changes and highlights the importance of the effect of anthropogenic halogenated GHGs on the global climate. Substantial time-series observations and statistical analyses of global lower-stratospheric temperature, troposphere–stratosphere temperature climatology, global surface air temperature, sea-ice extent and snow cover extent, together with observed data of stratospheric ozone depletion and stratospheric temperature reported in our recent papers [32, 33], combined with our theoretical calculations of lower-stratospheric temperatures and global mean surface temperatures, have provided solid and convincing evidence that ozone depletion and global climate changes are predominantly caused by halogen-containing ODSs and GHGs, respectively. There is no doubt that $CO_2$, $N_2O$ and $CH_4$ are important well-mixed GHGs that led to Earth's current ecological environment. However, the observed data strongly indicate that halogenated GHGs play a major role in causing global warming since 1950. Further, we also observe that both global lower-stratospheric temperature and snow cover extent in the NH or North America have stabilized over the past two and a half decades and, most remarkably, upper-stratospheric warming rather than cooling has been observed at high latitudes in the past decade. The observed data also show that a reversal in global warming would have occurred in around 2005 if there were no Arctic amplification of warming by sea-ice loss. This conclusion is striking, but it is consistent with previous observations [4, 5, 9, 10, 13, 16, 32, 33].

The observed results in Figures 1–8 and S3 do not agree with the climate models that assume $CO_2$ as the major culprit of warming. If the latter were correct, we would see continued large global surface warming and upper-stratospheric cooling, as widely believed [8]. Although the present study does not aim to rule out the $CO_2$-dominant warming climate models, the observations presented in this study strongly demonstrate that the observed global warming since the mid-1970s is dominantly caused by halogenated GHGs (mainly CFCs prior to 2000) and a warming reversal has begun. Therefore, this study calls for closer scrutiny of the assumptions made in and evidence for current climate models.

Since GMST is still around the peak, ice melting at the Arctic unavoidably remains and, in Antarctica, may increase with the recovery of the Antarctic ozone hole due to the increased greenhouse effect of recovering ozone. Thus, the warming amplification of sea-ice loss at the Arctic or Antarctic may continue to be significant for some years.

With the controls in global production and use of halogenated GHGs (CFCs, HCFCs, HFCs and PFCs) by international Agreements, including the highly successful and extremely important Montreal Protocol and its Amendments, however, it is very likely to see a gradual global reversal in GMST in the coming decades. Nevertheless, this expected reversal in global warming and the emerging shrinking of the ozone holes, including the freshly discovered tropical ozone hole, affecting approximately half of the world's population [33], will come true only with continued international efforts in phasing out all halogenated ozone-depleting substances and halogenated greenhouse gases. Therefore, the results of this study highlight the importance of such efforts from international governments and community. However, it is worth noting that the relevant international policies and political agenda, such as those attempted to make through the UN Climate Change Conferences (UNFCCCs), must be initiated and built on a solid foundation of correct understanding of the underlying mechanism of global climate change. This is of critical importance for humans, not only to reverse the climate change caused by anthropogenic drivers but to maintain a healthy economy and ecosystem around the globe.

**Supplementary Materials:** The following supporting information can be downloaded at: www.mdpi.com/xxx/s1, Figure S1:Contributions of natural ENSO and volcanic aerosols to observed GMSTs; Figure S2: Contributions of natural ENSO and volcanic aerosols to observed land surface air temperatures at NH extratropic (30°N-90° N); Figure S3:a-b: Time Series measured surface temperatures in North Europe (Sweden, Norway, Finland, UK, Ireland and Iceland) and North Asia including 11 countries (Afghanistan, Iran, Japan, Kazakhstan, N Korea, S Korea, Kyrgyzstan, Mongolia, Pakistan, Tajikistan, and Uzbekistan) plus 12 north and west provinces (Gansu, Hebei, Heilongjiang, Jilin, Liaoning, Nei Mongolia, Ningxia, Qinghai, Shaanxi, Shanxi, Xinjing, Xizang) of China.

**Funding:** This work is supported by the Natural Science and Engineering Research Council of Canada (Discovery grant # RGPIN-2017-05040).

**Data Availability Statement:** The data used for this study were obtained from the following sources: NOAA's Microwave Sounding Units (MSU) UAH and RSS datasets and Ratpac-B time-series dataset were obtained from NOAA (https://www.ncdc.noaa.gov/climate-monitoring/, accessed on 4 December 2021); radio occultation (RO) satellite datasets were obtained from the EUMETSAT's ROM SAF(https://www.romsaf.org/product_archive.php, accessed on 23 November 2021); zonal mean latitude–altitude distribution of $CF_2Cl_2$ (CFC-12) in 1992 was obtained from the NASA UARS (CLEAS) dataset (https://earthdata.nasa.gov/or https://uars.gsfc.nasa.gov/Public/Analysis/UARS/urap/home.html, accessed on 30 November 2021); Global land surface air temperature and GMST datasets were obtained from the UK Met Office (CRUTEM4.6, CRUTEM5.0 and HadCRUT4.6) (https://hadleyserver.metoffice.gov.uk/, accessed on 22 October 2021); land surface air temperatures of regions or individual countries were obtained from Berkeley Earth (http://berkeleyearth.org/data/, accessed on 20 November 2021), UK Met Office (HadCET) and NOAA (ClimDiv CONUS); sea ice extent, snow cover extent and ENSO datasets were obtained from NOAA; cosmic ray data were obtained from ref. [32]; updated concentrations of $CO_2$ and halogenated GHGs as well as calculated radiative forcings of $CO_2$ given in current climate models were obtained from the 2021 IPCC AR6 Report [8].

**Acknowledgments:** The author is greatly indebted to the Science Teams for making the data used for this study available.

**Conflicts of Interest:** The author declares no conflict of interest.